\documentclass[useAMS,usenatbib,onecolumn,usegraphicx]{mn2e}

\newcommand{\gsim}{\, \raisebox{-0.8ex}{$\stackrel{\textstyle >}{\sim}$ }}
\newcommand{\lsim}{\, \, \raisebox{-0.8ex}{$\stackrel{\textstyle <}{\sim}$ }}

\newcommand{\beq}{\begin{equation}}
\newcommand{\eeq}{\end{equation}}
\newcommand{\beqar}{\begin{eqnarray}}
\newcommand{\eeqar}{\end{eqnarray}}

\usepackage{ulem}
\usepackage{color}

\title[Stability of PNSs toward BH formation]  
{Stability of the protoneutron stars toward black hole formation} 
\author[H. Sotani \& K. Sumiyoshi]
{Hajime Sotani$^{1,2}$ \thanks{E-mail:sotani@yukawa.kyoto-u.ac.jp} and
Kohsuke Sumiyoshi$^{3}$
\\
$^1$Astrophysical Big Bang Laboratory, RIKEN, Saitama 351-0198, Japan\\
$^2$Interdisciplinary Theoretical \& Mathematical Science Program (iTHEMS), RIKEN, Saitama 351-0198, Japan\\
$^3$National Institute of Technology, Numazu College, Ooka 3600, Numazu, Shizuoka 410-8501, Japan}

\begin{document}
\maketitle
\label{firstpage}

\begin{abstract}
We examine the protoneutron star (PNS) stability in this study by solving the radial oscillation equations. For this purpose, we adopt the numerical results of massive PNS toward the black hole formation obtained by spherically symmetric numerical simulations for core-collapse supernova with general relativistic neutrino-radiation hydrodynamics. We find that the PNSs are basically stable in their evolution against the radial perturbations, while the PNS finally becomes unstable before the apparent horizon appears inside the PNS. We also examine the gravitational wave frequencies from PNS with the relativistic Cowling approximation. Then, we derive the empirical formula for the $f$-mode frequency, which weakly depends on the PNS models. This kind of universality tells us the PNS property, which is a combination of the PNS mass and radius in this study, once one would observe the $f$-mode gravitational waves.
\end{abstract}


\section{Introduction}
\label{sec:I}

Success of the direct detection of gravitational waves opens a new era, i.e., the gravitational wave astronomy, where gravitational waves can be used for extracting a celestial information together with the electromagnetic waves and neutrino signals. In fact, the gravitational waves from the binary black hole mergers and the binary neutron star merger(s) have already detected by the Laser Interferometer Gravitational wave Observatory (LIGO) \citep{LIGO} and Virgo \citep{Virgo}, which tells us the mass of the compact objects and constraint on the equation of state (EOS) for dense matter. In particular, when the gravitational wave has been detected from the binary neutron star merger, GW170817, the electromagnetic counterpart has also been observed \citep{GW6,EM}. Owing to this observation, it becomes obvious that the short gamma-ray burst comes from the binary neutron star merger. In addition, the Japanese gravitational wave detector, KAGRA \citep{KAGRA}, has finally joined the LIGO and Virgo collaboration network at the end of the third observing run, while the discussion for the third-generation detectors, such as Einstein Telescope and Cosmic Explorer \citep{punturo,CE}, has already been started. Thus, the number of gravitational wave events will increase more and more in the near future.

Supernova explosions must be another promising source of gravitational waves. The gravitational waves from core-collapse supernova have been mainly discussed with the numerical simulations of gravitational collapse of massive star, the bounce of central core, and subsequent accreting matter to the central compact object, i.e., protoneutron star (PNS), produced via supernova (e.g., \cite{Murphy09,MJM2013,Ott13,CDAF2013,Yakunin15,KKT2016,Andresen16,OC2018}). For instance, the three-dimensional (3D) numerical simulations for the core-collapse supernova tell us the existence of gravitational wave signals related to the oscillations of PNS (or core region) \citep{KKT2016,RMBVN19,Andresen19,Mezzacappa19}, where the frequency increases with time from a few hertz up to kilohertz after corebounce. This signal is originally considered to be the evidence of the Brunt-V\"{a}is\"{a}l\"{a} frequency at the PNS surface, i.e., the so-called surface gravity ($g$-) mode \citep{MJM2013,CDAF2013}, but it may be natural to consider that the signals come from the global oscillations of the PNSs because the Brunt-V\"{a}is\"{a}l\"{a} frequency is a local value and depends strongly on the definition of PNS radius. In fact, the gravitational wave signals appearing in the numerical simulations can be identified with the global oscillations, such as fundamental ($f$-mode) oscillations of PNS \citep{MRBV2018,SKTK2019,ST2020b,ST2020c}
 or the $g$-mode like oscillations on the region inside the shock radius \citep{TCPF2018,TCPOF2019,TCPOF2019b}. In addition to the ramp up signals, the 3D simulations tell us the existence of another gravitational wave signal with $\sim 100$ Hz, which seems to be associated with the standing accretion-shock instability \citep{KKT2016,Andresen16,OC2018,RMBVN19,Andresen19,Mezzacappa19}.

The numerical simulations must be a suitable technique for examining the dynamics of astronomical phenomena, while the perturbative approach is also powerful technique for extracting the physics behind the numerical results obtained by simulations. In particular, since a frequency from the astronomical objects strongly depend on their properties, one would inversely know the properties of objects through the observation of the frequency. This technique is known as asteroseismology, which is similar to seismology on the Earth and helioseismology on the Sun. As for (cold) neutron stars,  the crust properties are constrained by identified the quasiperiodic oscillations observed in the afterglow following magnetar giant flares with the crustal torsional oscillations (e.g., \cite{GNHL2011,SNIO2012,SNIO2013a,SNIO2013b,SIO2016,SIO2017,SIO2018,SIO2019}). Additionally, once the gravitational waves from compact objects will be detected, one could constrain the stellar mass, radius, and EOS of source objects (e.g., \cite{AK1996,AK1998,STM2001,SKH2004,TL2005,SYMT2011,PA2012,DGKK2013,Sotani20a,Sotani20b,Sotani21}).

On the other hand, the study on the PNS asteroseismology is relatively limited (e.g., \cite{FMP2003,Burgio2011,FKAO2015,ST2016,Camelio17,SKTK2017,MRBV2018,TCPF2018,TCPOF2019,TCPOF2019b,SKTK2019,ST2020a,ST2020b,ST2020c,Bizouard2021}), compared to that for cold neutron stars. Nevertheless, owing to the development of the numerical simulations of core-collapse supernova, the number of studies about the PNS asteroseismology gradually increases. Up to now, considering the PNS asteroseismology, two different approaches have been adopted. With one of two approaches, the PNS surface is defined with a specific surface density, which is typically $\sim 10^{11}$ g/cm$^3$, and the linear analysis is done inside the PNS \citep{MRBV2018,SKTK2019,ST2020a,ST2020b,ST2020c}. With this approach, there is uncertainty how to select the surface density, but at least the $f$- and $g_1$-mode frequencies for late phase after corebounce ($\sim 0.3$ sec.) are almost independent of the surface density \citep{MRBV2018,ST2020b}. As an advantage of this approach, since the boundary condition is the same as the standard asteroseismology, the classification of eigenmode is simple, i.e., the eigenmode is classified by counting the radial nodal number in the eigenfunction. Meanwhile, with the other approach, the linear analysis is done in the region inside the shock radius, i.e., there is no uncertainty for selection of the outer boundary \citep{TCPF2018,TCPOF2019,TCPOF2019b,Bizouard2021}. But then, the zero radial displacement condition is imposed at the outer boundary, which is different boundary condition from the standard asteroseismology. That is, the problem to solve with the second approach is completely different mathematical problem, where one has to reclassify the eigenmode with another definition. So, in this study we simply adopt the first approach and will discuss the stability of PNS and gravitational wave frequencies.

Most of the PNS asteroseismology among previous studies are considered with the PNSs produced by the  successful supernova, while in this study we focus on the accreting PNSs toward black hole formation due to the massive progenitor \citep{Sumiyoshi06,OConnor11,Pan18,Kuroda18}, i.e., the case for the failed supernova, as in  \cite{SS2019}. In the previous study \citep{SS2019}, we examine the gravitational wave frequency from the accreting PNSs and propose the possibility how to identify the PNS model with the increasing mass by using the simultaneous observation of the neutrino signal and gravitational waves from PNS (See \cite{Yokozawa15,Kuroda17,Takiwaki18,WS2019} for the simultaneous observations of the neutrinos and gravitational waves, for example).  However, the details of the gravitational waves from PNSs in the final phase just before the black hole formation are not discussed well in our previous study. In addition, the stability analysis on PNSs has been limited, 
although the presence of unstable quasi-radial modes has actually been discussed in the context of supernova  simulations in \cite{CDAF2013,TCPOF2019}, and \cite{TCPOF2019} computed the frequencies of the radial modes (among others) and estimated their gravitational wave emission depending on the deformation of the rotating PNS.
So, in this study we mainly examine the stability of PNS by considering the radial oscillation, as in cold neutron stars \citep{Chandrasekhar64}. Then, we also discuss the gravitational waves from the PNSs, especially focusing on the final phase just before the black hole formation, where we study the dependence of the gravitational wave frequencies on the surface density and derive a new fitting formula by adding new time periods from the previous analysis in \cite{SS2019}.

The paper is organized as follows. In section \ref{sec:PNSmodel}, we describe the setup of PNS models, which become background models for linear analysis, and their properties, especially focusing on the final stage in their evolution just before the black hole formation. In section \ref{sec:stability}, we examine the stability of protoneutron stars by considering the radial perturbations. In section \ref{sec:GW}, we report the eigenfrequencies of gravitational waves radiating from the PNS models. Finally, we summarize this study in section \ref{sec:Conclusion}.  Unless otherwise mentioned, we adopt geometric units in the following, $c=G=1$, where $c$ denotes the speed of light, and the metric signature is $(-,+,+,+)$.

\section{PNS Models}
\label{sec:PNSmodel}

In order to make a linear analysis, first we have to prepare the PNS models as a background. The PNS properties depend on not only the density and pressure profiles but also the distributions of temperature (or entropy per baryon) and electron fraction inside the PNS, while such profiles can be determined only via the numerical simulation of the core-collapse supernova explosion. In this study, as in \cite{SS2019}, we particularly adopt the profiles obtained via the numerical simulations performed by solving the general relativistic neutrino-radiation hydrodynamics under the spherical symmetry. In the simulations, hydrodynamics and neutrino transfer in general relativity are solved simultaneously \citep{Yamada97,Yamada99,Sumiyoshi05}. To describe the neutrino transfer, the Boltzmann equation is directly solved with the multiangle and multienegy neutrino distributions for four species, $\nu_e$, $\bar{\nu}_e$, $\nu_{\mu/\tau}$, and $\bar{\nu}_{\mu/\tau}$, i.e., we implement 6 species of neutrinos by assuming $\mu$-type and $\tau$-type (anti-)neutrinos have identical distributions. For the collision term associated with neutrino emission, absorption, and scattering with leptons, nucleons and nuclei handles, the basic neutrino reactions are adopted \citep{Sumiyoshi05,Bruenn85}. The metric adopted in the numerical code is given by
\begin{equation}
  ds^2 = -e^{2\Phi(t,m_b)}dt^2 + e^{2\Lambda(t,m_b)}dm_b^2 + r^2(t,m_b)(d\theta^2 + \sin^2\theta d\phi^2), \label{eq:metric1}
\end{equation}
where $t$ and $m_b$ denote the coordinate time and the baryon mass coordinate, respectively \citep{MS64}. In addition, $m_b$ is related to the circumference radius ($r$) via the baryon mass conservation, while the metric functions, $\Phi(t,m_b)$ and $\Lambda(t,m_b)$, are evolved together with hydrodynamical variables in the numerical simulations \citep{Yamada97}. The numerical simulations for core-collapse supernovae have been done with 255 grid points in the radial mass coordinate, 6 grid points in the neutrino angle, and 14 grid points in the neutrino energy. The rezoning of radial mesh is made during the simulations to resolve the accreting matter.  We remark that the radial grids of mass coordinate are nonuniformly arranged to cover not only the dense region inside the central object but also the region for accreting matter.

As in \cite{SS2019}, in order to examine the massive PNS and see the dependence on the progenitor models, we focus on two different progenitor models, i.e., a $40M_\odot$ star based on \cite{WW95} and a $50M_\odot$ star based on \cite{TUN07}, which are hereafter refereed to as W40 and T50, respectively (See \cite{Nakazato13,Sukhbold16,Horiuchi18} for the various models of progenitors and their consequences).  In addition, we adopt three different EOSs in this study, i.e., the Shen EOS \citep{Shen_EOS}, LS180, and LS220 \citep{LS_EOS}. The Shen EOS is based on the relativistic mean field theory with the TM1 nuclear interaction, while LS180 and LS220 are constructed with the compressible liquid drop model. In Table \ref{tab:EOS} we show the EOS parameters adopted in this study, i.e., the incompressibility, $K_0$, the symmetry energy, $S_0$, and the slope parameter of the nuclear symmetry energy, $L$, together with the maximum (gravitational) mass for the cold neutron stars constructed with the corresponding EOS \citep{Oertel17}. Even though the maximum mass with LS180 can not reach the $2M_\odot$ observations \citep{Demorest2010,Antoniadis2013,C20}, we also adopt LS180 to examine the dependence of the gravitational wave frequencies on the EOS stiffness \citep{Sumiyoshi04}. 
On the other hand, according to the terrestrial experimental constraints on the nuclear saturation parameters, $K_0$ is to be $ 230\pm 40$ MeV \citep{KM13}, while $L$ still seems to have a large uncertainty, such as $L=58.9\pm 16$ MeV \citep{Li19} and $L=106\pm 37$ MeV \citep{Reed21}. The three EOSs have been popularly used to explore the EOS dependence, although some of the EOS parameters are extreme in the experimental ranges. Anyway, one can see the details about the evolution of PNS toward the black hole formation and properties of neutrino emissions with these progenitor models and EOSs in \cite{Sumiyoshi07,Sumiyoshi08,Sumiyoshi09,Nakazato10}. With using the numerical data obtained in these simulations, we will prepare the PNS models and consider their oscillations in this study. As in \cite{SS2019}, we consider the PNS models provided by five sets of numerical simulations, i.e., W40 with Shen, LS180, and LS220 (three sets) and T50 with Shen and LS180 (two sets). To identify the PNS model, we hereafter label the model by combining the progenitor model with EOS, e.g., W40-Shen for the PNS model provided with W40 and Shen.

\begin{table}
\caption{EOS parameters adopted in this study, and the maximum (gravitational) mass for a cold neutron star constructed with the corresponding EOS.} 
\label{tab:EOS}
\begin {center}
\begin{tabular}{ccccc}
\hline\hline
EOS & $K_0$ (MeV) & $S_0$ (MeV) & $L$ (MeV) &  $M_{\rm max}/M_\odot$   \\
\hline
Shen & 281 & 36.9 & 110.8 & 2.2   \\
LS180 & 180 & 29.3 & 73.8 & 1.8   \\
LS220 & 220 & 29.3 & 73.8 &  2.0  \\
\hline \hline
\end{tabular}
\end {center}
\end{table}

\begin{figure}
\begin{center}
\includegraphics[scale=0.5]{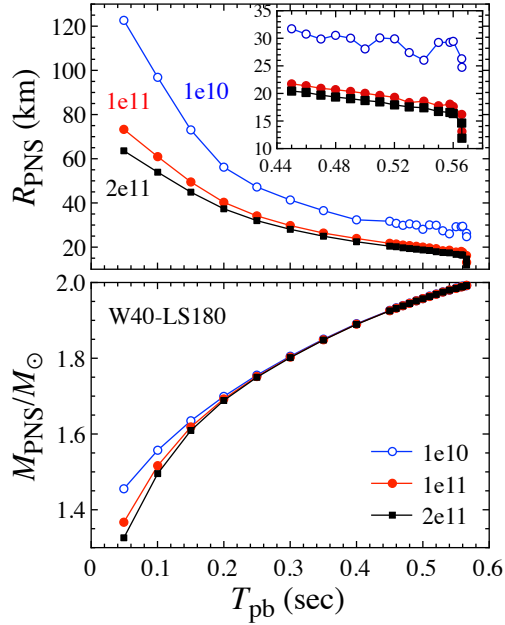} 
\end{center}
\caption{
Time evolution of the PNS radius (top) and gravitational mass (bottom) 
is shown as a function of $T_{\rm pb}$ for the model of W40-LS180 with $\rho_{\rm s} = 10^{10}$ (open-circles), $10^{11}$ (filled-circles), and $2\times 10^{11}$ g/cm$^3$ (filled-squars). In the top panel, an enlarged figure focusing on the final phase before the gravitational collapse is also added.
}
\label{fig:RMt-W40LS180}
\end{figure}

Unlike the cold neutron stars, the stellar surface of the PNS is ambiguous because the accreting matter exists even outside the PNS ``surface".  So, usually the PNS surface is identified with the position where the density becomes a specific surface density, $\rho_s$, e.g., typically $\rho_s\simeq 10^{11}$ g/cm$^3$. In Fig. \ref{fig:RMt-W40LS180}, we show the $\rho_s$-dependence of the PNS radius (top panel) and gravitational mass (bottom panel) 
for the model of W40-LS180 with $\rho_s=10^{10}$, $10^{11}$, and $2\times 10^{11}$ g/cm$^3$, where $T_{\rm pb}$ denotes a time after the core bounce in the unit of seconds, i.e., the moment at the core bounce corresponds to $T_{\rm pb}=0$. From this figure, one can observe a wiggling behavior in the PNS radius with lower surface density. This is because the numerical simulations have been done with the Lagrangian coordinate in baryon mass instead of Eulerian radial coordinate.  Resolving the gradually accreting matter in the moving baryon mass grids is well done by the rezoning method, however, determining the radial position is not as easy as in the fixed radial grids.  As a result, the number of grid points in the low density region is limited, i.e., the resolution is not enough to get the smooth radius evolution with lower surface density. 
Nevertheless, since the radius evolution for the PNS model with $\rho_s=2\times 10^{11}$ g/cm$^3$ is relatively smooth even in the final phase, we adopt $\rho_s=2\times 10^{11}$ g/cm$^3$ in this study and discuss the stability and gravitational wave frequencies on such PNS models. We remark that the PNS mass seems to be almost independent of the selection of surface density.

\begin{figure}
\begin{center}
\includegraphics[scale=0.5]{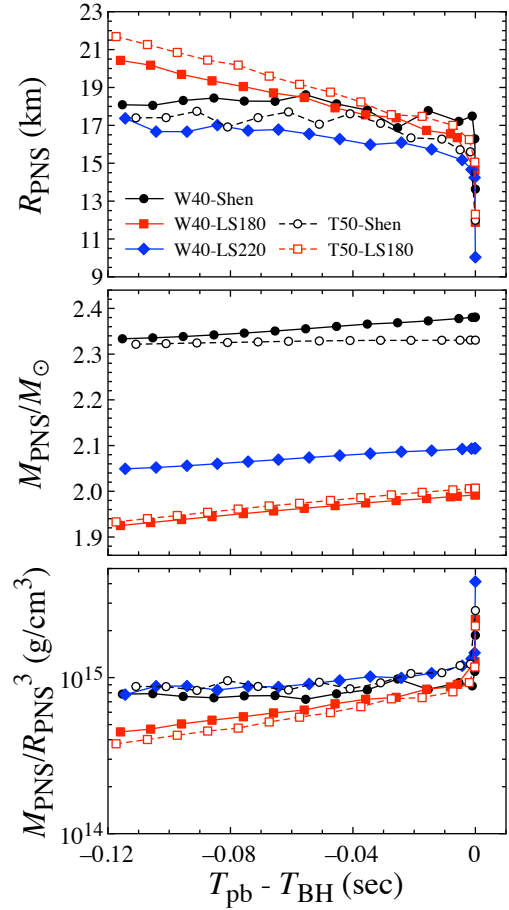} 
\end{center}
\caption{
Time evolution of the PNS radius (top), gravitational mass (middle), and average density (bottom) for the final phase before the gravitational collapse is shown as a function of $T_{\rm pb} - T_{\rm BH}$ for various models with $\rho_{\rm s} = 2\times 10^{11}$ g/cm$^3$.
}
\label{fig:Evo-RMave}
\end{figure}

In Fig. \ref{fig:Evo-RMave}, we show the time evolution of the PNS radius (top panel), gravitational mass (middle panel), and average density defined by $M_{\rm PNS}/R_{\rm PNS}^3$ (bottom panel) as a function of $T_{\rm pb}-T_{\rm BH}$ for various models with $\rho_s=2\times 10^{11}$ g/cm$^3$ \footnote{The label on the $y$ axis in the right panel of Fig. 3 in \cite{SS2019} is typo, which should be the stellar average density, $M_{\rm PNS}/R_{\rm PNS}^3$, instead of the square root of it, $(M_{\rm PNS}/R_{\rm PNS}^3)^{1/2}$}, where $T_{\rm BH}$ denotes the black hole formation time with $T_{\rm pb}$ and the concrete values are listed in Table \ref{tab:TBH}. In the current study, we define the black hole formation time by finding the apparent horizon, i.e., the PNS models adopted in this study cover the evolution up to the final collapse to the black hole formation\footnote{The simulations have been performed even a little after the formation of apparent horizon in most of simulations except for T50-Shen.}.  The apparent horizon is formed in the snapshot at the final time step except for the case of T50-Shen.  We note that the values of $T_{\rm BH}$ shown in Table \ref{tab:TBH} are a little larger than those shown in Table I in  \cite{SS2019}, where the final snapshot is just before the collapse to the black hole.  Toward the moment of the black hole formation, $T_{\rm BH}$, the PNS radius suddenly decreases as shown in Figs. \ref{fig:RMt-W40LS180} and \ref{fig:Evo-RMave}. From Table \ref{tab:TBH}, one can see that the black hole formation time is shorter for the PNS models with softer EOS. In Fig. \ref{fig:Evo-RMave}, the time variation of the PNS radius and the average density change faster with softer EOS. 
In addition, we remark that the maximum gravitational mass for PNSs is $\sim 10 \%$ larger than that for cold neutron stars, which comes from the thermal effect and the contribution of lepton, such as an electron and neutrino.

\begin{table}
\centering
\caption{$T_{\rm BH}$ and $T_{f_{\xi}=0}$ for various models, where $T_{\rm BH}$ and $T_{f_{\xi}=0}$ are respectively the black hole formation time, i.e., the time when the apparent horizon appears inside the PNS, and the time when the fundamental frequency of the radial oscillation becomes zero.} 
\begin{tabular}{cccc}
\hline\hline
EOS & Model &  $T_{\rm BH}$ (sec) &  $T_{f_{\xi}=0}$ (sec)  \\
\hline
Shen   & W40  &  1.345  &  1.345   \\
            & T50   &  1.511  &  1.509 \\
LS180 & W40  &  0.566  &  0.561  \\
            & T50  &   0.507  &  0.506  \\
LS220 & W40  &  0.784  &  0.782  \\
\hline \hline
\end{tabular}
\label{tab:TBH}
\end{table}

Finally, as in the previous studies (e.g., \cite{ST2016,MRBV2018,SKTK2019,TCPF2018,TCPOF2019,SS2019}), we assume that the PNS model are in a static equilibrium at each time step and prepare the PNS model as a background model for considering the linear analysis in this study. In this case, the metric is given with the spherical coordinate as
\begin{equation}
  ds^2 = -e^{2\Phi}dt^2 + e^{2\Lambda}dr^2 + r^2\left(d\theta^2 + \sin^2 d\phi^2\right).
\end{equation}
Now, $\Phi$ and $\Lambda$ are functions only of $r$ and $e^{2\Lambda}$ is directly associated with the mass function $m(r)$ via $e^{-2\Lambda}=1-2m/r$.

\section{Stability of Protoneutron stars}
\label{sec:stability}

As seen in the previous section, the mass of the protoneutron star increases with the accretion, which eventually approaches the maximum mass allowed with the adopted EOSs. Then, the protoneutron star would gravitationally collapse to a black hole. The moment when the protoneutron star approaches its maximum mass corresponds to the onset of the instability. In order to determine the onset of instability in the evolution of the protoneutron stars, we make a linear analysis with the radial perturbation on the protoneutron star models at each time step after core bounce. For this purpose, one can derive the perturbation equations as
\begin{eqnarray}
  \frac{d\xi}{dr} &=& -\left[\frac{3}{r} + \frac{p'}{p+\varepsilon}\right]\xi - \frac{\eta}{r\Gamma},
       \label{eq:dxi} \\
  \frac{d\eta}{dr} &=& \left[r(p+\varepsilon) e^{2\Lambda}\left(\frac{\omega^2}{p} e^{-2\Phi} - 8\pi \right) 
      - \frac{4p'}{p} + \frac{r(p')^2}{p(p+\varepsilon)}\right]\xi
      - \left[\frac{\varepsilon p'}{p(p+\varepsilon)} + 4\pi  r(p+\varepsilon)e^{2\Lambda}\right]\eta, \label{eq:deta}
\end{eqnarray}
where $p$, $\varepsilon$, and $\Gamma$ denote the pressure, energy density, and adiabatic index for the background protoneutron star models, while $\xi$ and $\eta$ are perturbative variables given by $\xi\equiv \Delta r/r$ and $\eta\equiv \Delta p/p$ with the radial displacement, $\Delta r$, and the Lagrangian perturbation of pressure, $\Delta p$ \citep{Chandrasekhar64,Chanmugan77,GHZ97} 
\footnote{The perturbation equations for radial oscillations considered in this study are derived without the Cowling approximation, while it is assumed only in the calculation of the non-radial oscillations discussed in the section \ref{sec:GW}. In fact, if one assumes  the Cowling approximation for the radial oscillations, one can not discuss the onset of a black hole formation \citep{TCPF2018}.}. 
The prime in the equations denotes the radial derivative and the adiabatic index is given by 
\begin{equation}
  \Gamma \equiv \left(\frac{\partial \ln p}{\partial \ln n_{\rm b}}\right)_s = \frac{p+\varepsilon}{p}c_s^2, \label{eq:Gamma}
\end{equation}
where $n_{\rm b}$, $s$, and $c_s$ denote the baryon number density, entropy per baryon, and sound velocity, respectively. We remark that one can derive the Sturm-Liouville type second order differential equation with respect to $\xi$ from Eqs. (\ref{eq:dxi}) and (\ref{eq:deta}). To solve the eigenvalue problem with respect to the eigenvalue $\omega^2$, one should impose the appropriate boundary conditions. The boundary condition at the stellar surface comes from the condition to remove the singularity in Eq. (\ref{eq:deta}) \citep{Chanmugan77}, i.e.,
\begin{equation}
 \eta = -\left[\left(\frac{\omega^2R_{\rm PNS}^3}{M_{\rm PNS}} + \frac{M_{\rm PNS}}{R_{\rm PNS}}\right)
             \left(1-\frac{2M_{\rm PNS}}{R_{\rm PNS}}\right)^{-1}+ 4\right]\xi, \label{eq:BC1}
\end{equation}
while the boundary condition at the center is the regularity condition, i.e., 
\begin{equation}
  3\Gamma \xi + \eta = 0. \label{eq:BC0}
\end{equation}
In addition, as normalization of the eigenfunction, we set $\xi=1$ at the stellar center. Then, with the resultant eigenvalue $\omega^2$, the frequency of radial oscillations are given by
\begin{equation}
  f_{\xi} = {\rm sgn}(\omega^2)\sqrt{|\omega^2|}/2\pi, \label{eq:fxi}
\end{equation}
where the system is unstable when $\omega^2$ becomes negative.

\begin{figure*}
\begin{center}
\includegraphics[scale=0.5]{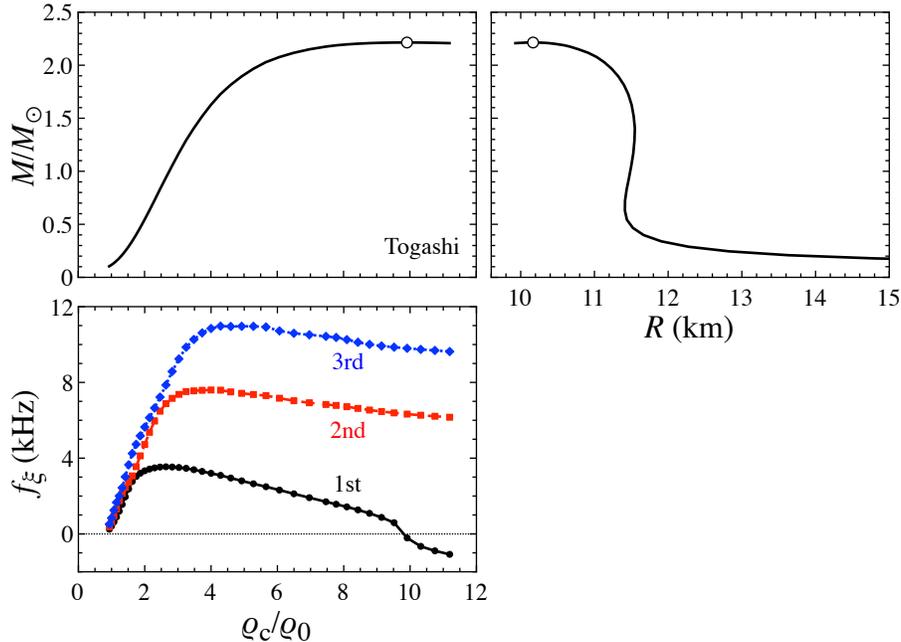} 
\end{center}
\caption{
For the neutron stars constructed with Togashi EOS \citep{Togashi-EOS}, the mass of stars is shown as a function of the central density, $\rho_c$, normalized by the nuclear saturation density, $\rho_0$, in the left-top panel and as a function of the stellar radius in the right-top panel, while the corresponding frequencies of the lowest three radial oscillations are shown as a function of $\rho_c/\rho_0$ in left-bottom panel. The open-circle denotes the stellar model with the maximum mass. It is obvious that the frequency of the lowest (1st) radial oscillation becomes negative for the stellar model constructed with the central density, which is more than that for the neutron star with the maximum mass. We remark that the central density in this figure is not a rest-mass density, but an energy density.
}
\label{fig:Cold-Togsahi}
\end{figure*}

For the case of cold neutron stars, it is well known that the stellar models constructed with the central density, which is more than that for the neutron star model with the maximum mass, are unstable. As an example (and as a test of our code for the eigenvalue problem with respect to the radial oscillations), we show the results for the neutron star models constructed with Togashi EOS \citep{Togashi-EOS} in Fig. \ref{fig:Cold-Togsahi}, where top panels correspond to the equilibrium models, i.e., the stellar mass is shown as a function of the central density, $\rho_c$, normalized by the nuclear saturation density, $\rho_0$, in the left-top panel and as a function of the stellar radius in the right-top panel. The open-circle denotes the stellar model with the maximum mass. In the bottom-left panel, we show the frequencies of the lowest three radial oscillations as a function of $\rho_c/\rho_0$. From this figure, it is obvious that the frequency of the lowest radial oscillation becomes negative for the stellar models with the central density, which is larger than that for the stellar model with the maximum mass, i.e., the corresponding stellar models are unstable.

\begin{figure*}
\begin{center}
\includegraphics[scale=0.5]{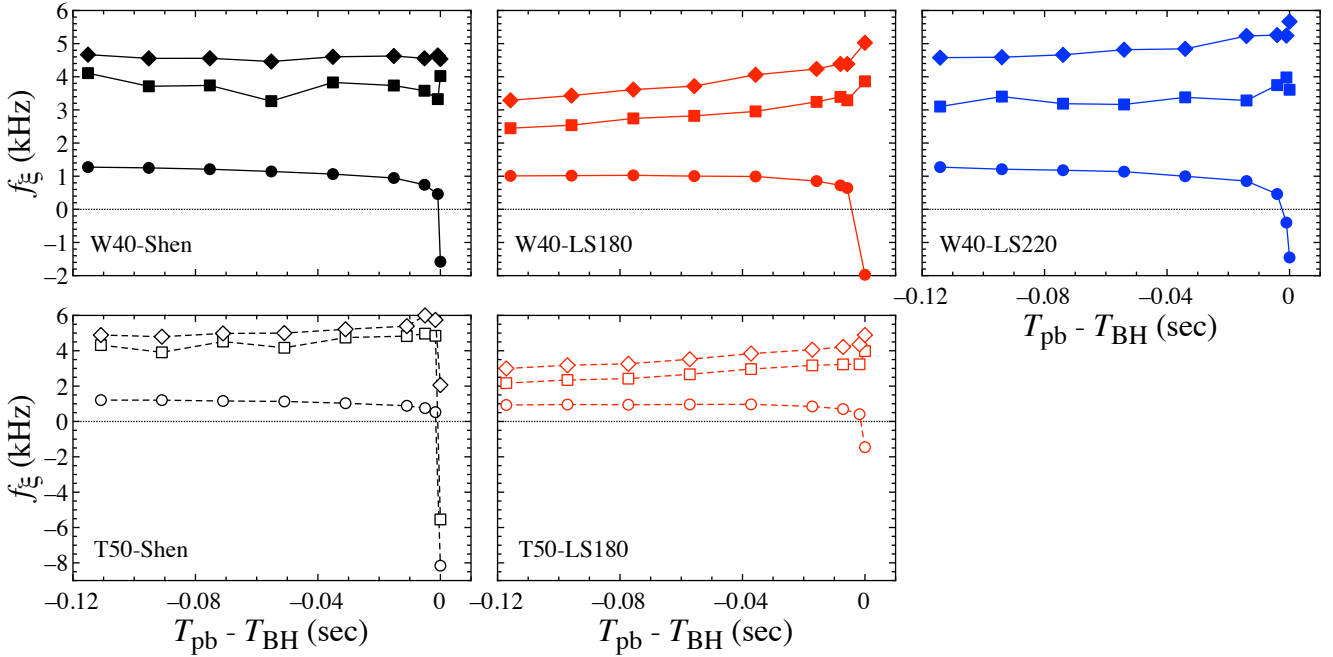} 
\end{center}
\caption{
The frequencies of the lowest three radial oscillations are shown as a function of $T_{\rm pb} - T_{\rm BH}$, where the top and bottom panels correspond to the PNS models with W40 and T50, while the left, middle and right panels correspond to the PNS models with Shen, LS180, and LS220, respectively. 
}
\label{fig:radial}
\end{figure*}

Now, we consider the case of PNSs. In Fig. \ref{fig:radial}, we show the frequencies of the lowest three radial oscillations on the various PNS models for the final phase just before the black hole formation. From this figure, one can obviously see that, for any PNS models considered in this study, the lowest frequency is basically positive, i.e., PNS is stable, but it becomes negative at the last moment. That is, the PNS model for which the lowest frequency is zero is the PNS model with the maximum mass allowed with the adopted EOS, where the corresponding time is shown in Table \ref{tab:TBH}. Since one can see that $T_{\rm BH} > T_{f_{\xi}=0}$ from Table  \ref{tab:TBH} and Fig. \ref{fig:radial}, we find that the PNS becomes unstable before the apparent horizon appears inside the PNS. In addition, we find that the lowest frequency monotonically decreases with time at least in the phase considered in this study, while the second and third lowest frequencies for the models with soft EOSs increase. Incidentally, the second and third lowest frequencies become negative at the end as for the model of T50-Shen.

\begin{figure*}
\begin{center}
\includegraphics[scale=0.5]{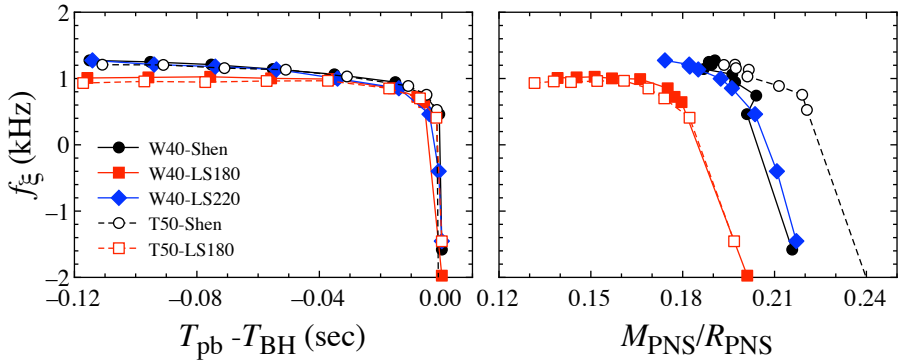} 
\end{center}
\caption{
For various PNS models, the frequencies of the 1st radial oscillations for the final phase are shown as a function of $T_{\rm pb} - T_{\rm BH}$ (left panel) and $M_{\rm PNS}/R_{\rm PNS}$ (right panel).
}
\label{fig:fxi}
\end{figure*}

Furthermore, we focus on the lowest frequency in Fig. \ref{fig:fxi}, where the lowest frequency is shown as a function of $T_{\rm pb}-T_{\rm BH}$ (left panel) and the PNS compactness (right panel). From the left panel, we see that the time evolution of the lowest frequency (especially in the last $\sim 40$ msec) weakly depend on the PNS model, although the second and third lowest frequencies depend on the PNS model, as mentioned above. We found that, from the right panel, the gradient of the lowest frequency with respect to the PNS compactness seems to be almost independent of the PNS models at the last moment, although as shown in Fig. \ref{fig:Evo-RMave} the PNS properties in the final phase depend on PNS models. In practice, the radial oscillation itself is not associated with the gravitational radiations, because radial oscillations of a spherical background do not induce a quadrupolar deformation and hence no gravitational waves, but one can observe the aspect of the radial oscillations in the gravitational waves through the nonlinear coupling between the radial and nonradial oscillations \citep{PSN07}.
On the other hand, apart from non-linear coupling, deformations due to stellar rotation can also make the radial oscillations visible \citep{CDAF2013,TCPOF2019b}. In presence of rotation, the radial oscillations become quasi-radial, however, for sufficiently slow rotation their frequencies should be close to the radial modes computed here.

\section{Protoneutron star asteroseismology}
\label{sec:GW}

In this study, we also examine the gravitational wave frequency from the PNSs in the final phase just before the black hole formation. For this purpose, as in \cite{SS2019} we simply adopt the relativistic Cowling approximation, i.e, the metric perturbations are neglected during the fluid oscillations, where one can derive the perturbation equations by linearizing the energy-momentum conservation law. Then, by adopting the appropriate boundary conditions at the stellar center and the radius, the problem to solve becomes an eigenvalue problem with respect to the eigenvalue $\omega$, with which the gravitational wave frequency, $f$, is determined via $f=\omega/(2\pi)$. The perturbation equations and boundary conditions are the same as in \cite{SKTK2019}. That is, we make a linear analysis in a similar way to the previous study \citep{SS2019}, but we especially focus on the final phase toward the black hole formation in this study. In this study, we only consider the $\ell=2$ mode oscillations, which are considered to be energetically dominant in gravitational wave radiation.

First, we examine how the gravitational wave frequencies from the PNSs depend on the PNS surface density. This dependence has already been discussed in \cite{MRBV2018,ST2020b} for the PNSs produced by the successful supernova explosions, where the $f$- and $g_1$-mode frequencies are independent of the selection of the surface density after the avoided crossing between the both modes, while the other modes, especially the $p_i$-modes with $i\ge 2$, depend on the surface density. This behavior may be understood by considering the radial profile of pulsation energy density (or eigenfunction) \citep{ST2020b}. That is, the $f$- and $g_1$-modes significantly oscillate inside the star, while the other modes oscillate not only inside the star but also in the region closed to the PNS surface. 

\begin{figure}
\begin{center}
\includegraphics[scale=0.5]{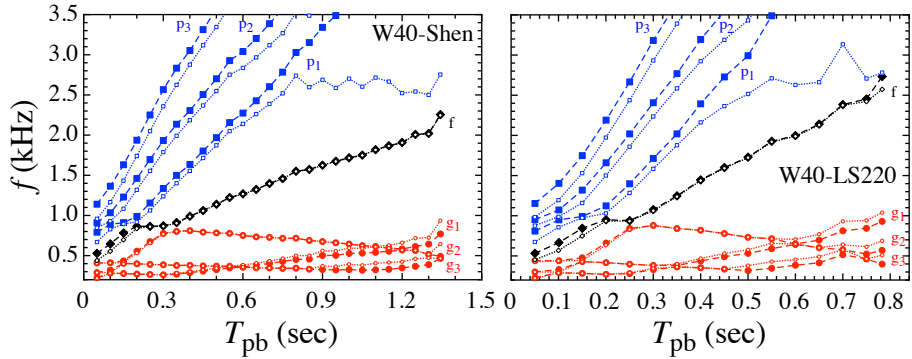} 
\end{center}
\caption{
Comparison between the eigenfrequencies of the $f$-, $p_i$-, and $g_i$-modes with $i=1-3$ for the PNSs models with $\rho_s=2.0\times 10^{11}$ g/cm$^3$ (filled-mark with dashed line) and those with $\rho_s=10^{11}$ g/cm$^3$ (open-mark with dashed line) for W40-Shen (left panel) and W40-LS220 (right panel). Although we do not show, the additional eigenfrequencies exist above the $p_3$- and below the $g_3$-modes, which are respectively the $p_i$- and the $g_i$-modes with $i\ge 4$.
}
\label{fig:dep-rhos}
\end{figure}

In the case for the PNSs in the failed supernovae considered in this study, we show the gravitational wave frequencies of the $f$-, $p_i$-, and $g_i$-modes with $i=1-3$ in Fig. \ref{fig:dep-rhos} for the PNS models of W40-Shen (left panel) and W40-LS220 (right panel) with $\rho_s=10^{11}$ (open-marks) and $2\times 10^{11}$ g/cm$^3$ (filled mark). From this figure, we confirm that the $f$- and $g_1$-mode frequencies are basically independent of the selection of the surface density even for the failed supernovae after the avoided crossing between the $f$- and $g_1$-modes.  We also find that the dependence of the surface density can be seen in the $g_1$-mode frequency in the final phase just before the black hole formation, which is the period after the avoided crossing between the $g_1$- and $g_2$-modes. In addition, as in the previous studies for the PNSs produced via the successful supernova explosions, the other modes seem to depend on the selection of surface density, where we find that the $p_1$-mode frequency significantly depends on the surface density in the late phase after the core bounce. We remark that a wiggling in the $p_1$-mode frequency with $\rho_s=10^{11}$ g/cm$^3$ may come from the fact that the numerical resolution is not so high in a lower density region.

\begin{figure}
\begin{center}
\includegraphics[scale=0.5]{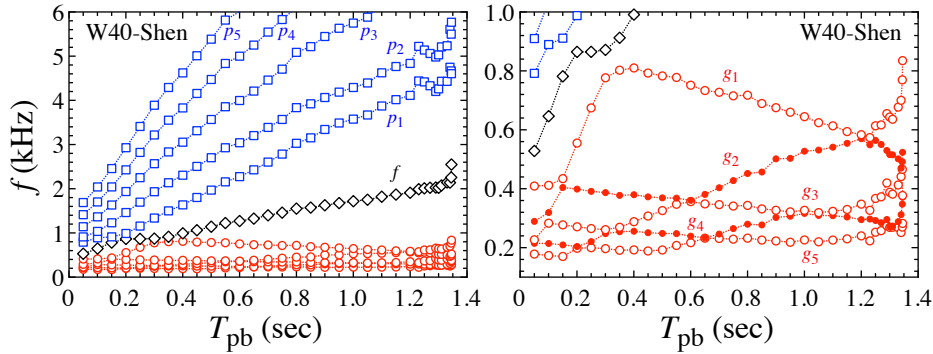} 
\end{center}
\caption{
Time evolution of gravitational wave frequencies from the PNS model of W40-Shen with $\rho_s=2\times 10^{11}$ g/cm$^3$, where the $f$-, $p_i$-, and $g_i$-modes with $i=1-5$ are shown up to the black hole formation. The right panel is just an enlarged view of the left panel. In order to focus on the PNS models in the final phase just before the black hole formation.  Note that we adopt a small time interval for the snapshots for the final phase.  
}
\label{fig:W40Shen}
\end{figure}

Next, we focus on the gravitational wave frequencies from the PNS in the final phase just before the black hole formation. In Fig. \ref{fig:W40Shen} we show the gravitational wave frequencies from the PNS model of W40-Shen with $\rho_s=2\times 10^{11}$ g/cm$^3$.  Note that we adopt a small time interval for the PNS models in the final phase to study in detail the black hole formation.  From this figure, one can observe that the gravitational wave frequencies steeply increase at the final moment when the PNS collapses to a black hole. This behavior can be seen not only in the PNS model of W40-Shen but also in the other PNS models as shown in appendix \ref{sec:appendix_1}. Thus, this steep increase in the gravitational wave frequencies would tell us the moment of the black hole formation. This feature is revealed in the current study, more clearly than \cite{SS2019} where the analysis is made only before the final collapse.

\begin{figure}
\begin{center}
\includegraphics[scale=0.5]{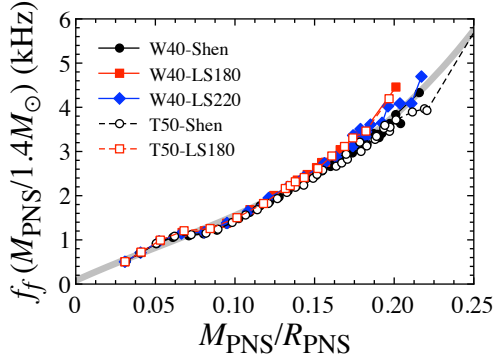} 
\end{center}
\caption{
The $f$-mode frequency multiplied by the normalized PNS mass is show as a function of the PNS compactness for various PNS models up to the black hole formation. The thick-solid line denote the fitting formula given by Eq. (\ref{eq:ffM}).
}
\label{fig:ffM}
\end{figure}

We explore the empirical formulae for the evolution of frequency of gravitational waves. 
In \cite{SS2019}, we just derive a fitting formula for the $f$-mode frequency as a function of the square root of the PNS average density according to \cite{AK1996,AK1998}, because the $f$-mode oscillation is a kind of an acoustic oscillation, which could be associated with the density of the object.  On the other hand, \cite{TL2005,Sotani21} shows another possibility for fitting of the $f$-mode frequency, which seems to be better than that in \cite{AK1996,AK1998}. In fact, as in \cite{TL2005,Sotani21}, we find that the $f$-mode frequency multiplied by the PNS mass, $f_f(M/1.4M_\odot)$, can be expressed as a function of the PNS compactness, ${\cal C}\equiv M_{\rm PNS}/R_{\rm PNS}$, almost independently of the PNS models even for the last PNS model in the evolution considered in this study, such as
\begin{equation}
  f_f \left(\frac{M_{\rm PNS}}{1.4M_\odot}\right)  {\rm (kHz)} = 0.08087 + 3.262\left(\frac{\cal C}{0.2}\right) 
        - 1.740\left(\frac{\cal C}{0.2}\right)^2 + 2.204\left(\frac{\cal C}{0.2}\right)^3.
        \label{eq:ffM}
\end{equation}
In Fig. \ref{fig:ffM} we show $f_f(M/1.4M_\odot)$ as a function of ${\cal C}$ for various PNS models, where the thick-solid line denotes the fitting formula given by Eq. (\ref{eq:ffM}). In this figure, the PNSs evolve from the bottom left to the top right. With this empirical formula, one can obtain a constraint on the relation between the PNS mass and radius at each time step by observing the $f$-mode gravitational wave from the PNS \footnote{The similar universal relation has been also derived in \cite{TCPOF2019b}, where the mode classification is different from ours because the position of outer boundary and the boundary condition are different from us. Additionally, in \cite{Bizouard2021} it is also shown that PNS asteroseismology using gravitational waves is possible in practice.}. 
We remark that what is called here $f$-mode is also interpreted as a surface $g$-mode \citep{MJM2013,CDAF2013} or a $g$-mode like oscillations by adopting the outer boundary at the shock radius \citep{TCPOF2019,TCPOF2019b}, where the corresponding frequencies can be expressed well with the PNS surface gravity,  $M_{\rm PNS}/R_{\rm PNS}^2$.

\begin{figure}
\begin{center}
\includegraphics[scale=0.5]{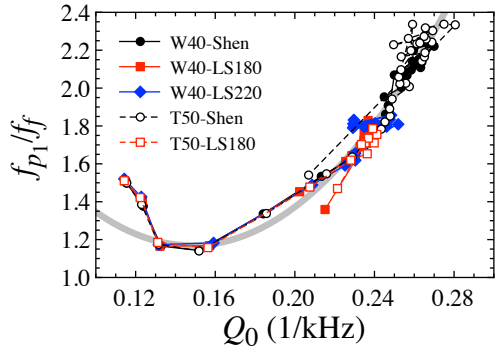} 
\end{center}
\caption{
For various PNS models up to the black hole formation, the ratio of the $p_1$-mode frequency to the $f$-mode frequency, $f_{p_1}/f_f$, is shown as a function of $Q_0$ defined by Eq. (\ref{eq:Q0}). The thick-solid line denotes the fitting formula given by Eq. (\ref{eq:fp1ff}), which is derived with using the data for $Q_0\gsim 0.13$ kHz$^{-1}$.
}
\label{fig:fp1_ff}
\end{figure}

In addition, it is pointed out that the ratio of the $p_1$-mode frequency to the $f$-mode frequency, $f_{p_1}/f_f$, can be characterized by a new variable, $Q_0$, defined in \cite{Camelio17} as
\begin{equation}
  Q_0 = \frac{1}{f_f} \left(\frac{M_{\rm PNS}}{1.4M_\odot}\right)^{1/2}\left(\frac{R_{\rm PNS}}{10\ {\rm km}}\right)^{-3/2}. \label{eq:Q0}
\end{equation}
In \cite{SS2019}, we could not find any good relation between $f_{p_1}/f_f$ and $Q_0$ \footnote{Explicitly speaking, we could not find any relation between $f_f/f_{p_1}$ (instead of  $f_{p_1}/f_f$) and $Q_0$ in \cite{SS2019}.}, where the PNS surface density is chosen to be $\rho_s=10^{11}$ g/cm$^3$. On the other hand, in this study with $\rho_s=2\times 10^{11}$ g/cm$^3$, we can get a good relation between $f_{p_1}/f_f$ and $Q_0$. In Fig. \ref{fig:fp1_ff}, $f_{p_1}/f_f$ is plotted as a function of $Q_0$ for various PNS models. In this figure, the PNSs evolve from left to right for the early phase, but their evolutionary direction changes from right to left for the later phase. From this figure, one can observe that $f_{p_1}/f_f$ is almost independent of the PNS models for $Q_0\lsim 0.2$ kHz$^{-1}$ and weakly depends on the PNS models for $Q_0\gsim 0.2$ kHz$^{-1}$. With using the data for $Q_0\gsim 0.13$ kHz$^{-1}$, we can obtain the fitting formula such as
\begin{equation}
  f_{p_1}/f_f = 2.845 - 22.608\left(\frac{Q_0}{1 {\rm /kHz}}\right) + 76.226\left(\frac{Q_0}{1 {\rm /kHz}}\right)^2, 
       \label{eq:fp1ff}
\end{equation}
with which the expected value of $f_{p_1}/f_f$ is plotted with the thick-solid line in Fig. \ref{fig:fp1_ff}. As mentioned above, since the $p_1$-mode frequency depends on the selection of the PNS surface density, it is still uncertain  how universal this fitting formula is. But, if this type of fitting formula truly exists, the simultaneous observation of the $f$- and $p_1$-mode gravitational waves tells us another constraint on the relation between the PNS mass and radius at each time step.
The dependence of the $p_1$-mode frequency on the PNS surface density comes from the fact that the pulsation energy density for the $p_1$-mode oscillates not only inside the PNS but also in the vicinity of the PNS surface \citep{ST2020b}. To determine the $p_1$-mode frequency with a realistic situation, one may have to impose the outer boundary condition at the shock radius, instead of the PNS surface, as in  \cite{TCPF2018,TCPOF2019,TCPOF2019b}.

\begin{figure}
\begin{center}
\includegraphics[scale=0.5]{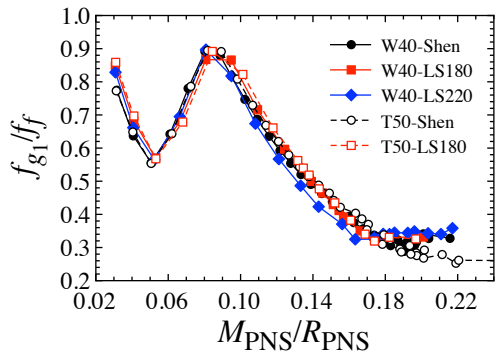} 
\end{center}
\caption{
For various PNS models up to the black hole formation, the ratio of the $g_1$-mode frequency to the $f$-mode frequency, $f_{g_1}/f_f$, is shown as a function of the PNS compactness.
}
\label{fig:fg1_ff}
\end{figure}

Finally, we also check the dependence of the ratio of the $g_1$-mode frequency to the $f$-mode frequency, $f_{g_1}/f_f$, on the PNS compactness. In \cite{SS2019}, it is already shown that $f_{g_1}/f_f$ can be characterized by the PNS compactness almost independently of the PNS models. We found that this universality with the frequencies calculated in this study holds up to the final moment of black hole formation, as shown in Fig. \ref{fig:fg1_ff}. Again, the simultaneous observation of the $f$- and $g_1$-mode gravitational waves tells us a constraint on the PNS compactness at each time step.

\section{Conclusion}
\label{sec:Conclusion}

We examine the PNS stability and gravitational wave frequencies from the massive PNS produced by failed supernovae. In order to prepare the PNS model, we adopt the numerical data obtained by the spherically symmetric numerical simulation performed by solving the general relativistic neutrino-radiation hydrodynamics. Then, as in the previous studies, we assume that the PNS models are static equilibrium at each time step. On such a PNS model, first we examine the PNS stability by solving the radial oscillations. We show that most of PNS models in their  evolution are stable against the radial perturbations, i.e., all the frequencies of the radial oscillations are positive, while the PNS model finally becomes unstable before the apparent horizon appears inside the PNS. We also show that the time evolution of the lowest frequency of the radial oscillation for the last $\sim 40$ msec weakly depend on the PNS model. In addition, we examine the gravitational wave frequencies from the PNS models by adopting the relativistic Cowling approximation. In this study we especially focus on the final phase in this study, although how to analysis is the same as in the previous study. We find that the frequencies steeply increase at the final moment when a black hole forms. So, such a signal in gravitational wave frequencies would tell us the moment of black hole formation. We also show that the $f$-mode frequency multiplied by the PNS mass can be characterized by the PNS compactness almost independently of the PNS models and derive the empirical formula for the $f$-mode frequency. Moreover, we find that the ratio of the $p_1$-mode to the $f$-mode weakly can be characterized by the variable, which is defined as the square root of the normalized PNS average density divided by the $f$-mode frequency, while the ratio of the $g_1$-mode to the $f$-mode can be characterized by the PNS compactness. So, via the simultaneous observation of the $f$- and $p_1$-mode (or the $f$- and $g_1$-mode) gravitational wave, one would constrain the PNS mass and radius at each time step, which helps us to understand the dense matter physics. In this study, as a first step, we simply assume that the PNS models are static equilibrium at each time step even for the final phase, which could be highly dynamical phase. Since this static assumption may not be valid on such a dynamical phase, we will study the PNS oscillations without the static assumption somewhere in the future. In fact, by comparing the dynamical timescale of the system defined as $\tau_{\rm dyn} = (R_{\rm PNS}^3/M_{\rm PNS})^{1/2}$ with the typical timescale of the $f$-mode oscillation, i.e., $1/f_f$, we find that the value of $f_f \tau_{\rm dyn}$ becomes $\sim 0.3$ in the final dynamical phase.

\section*{Acknowledgements}
This work is supported in part by Japan Society for the Promotion of Science (JSPS) KAKENHI Grant Numbers JP17H06357, JP17H06365, JP18H05236, JP19KK0354, JP19K03837, JP20H01905, JP20H04753, and JP21H01088
and 
by Pioneering Program of RIKEN for Evolution of Matter in the Universe (r-EMU).
For providing high performance computing resources, 
Computing Research Center, KEK, 
JLDG on SINET4 of NII, 
Research Center for Nuclear Physics, Osaka University, 
Yukawa Institute of Theoretical Physics, Kyoto University, 
Nagoya University, Hokkaido University, 
and 
Information Technology Center, University of Tokyo are acknowledged. 
%
This work was partly supported by 
MEXT as ``Program for Promoting Researches on the Supercomputer Fugaku"
(Toward a unified view of the universe: from large scale structures to planets)
and
the Particle, Nuclear and Astro Physics Simulation Program (Nos. 2020-004, 2021-004) of Institute of Particle and Nuclear Studies, High Energy Accelerator Research Organization (KEK).

\section*{Data availability}
The data underlying this article will be shared on reasonable request to the corresponding author.

\appendix
\section{Time evolution of gravitational wave frequencies from various PNS models}   
\label{sec:appendix_1}

\begin{figure}
\begin{center}
\includegraphics[scale=0.5]{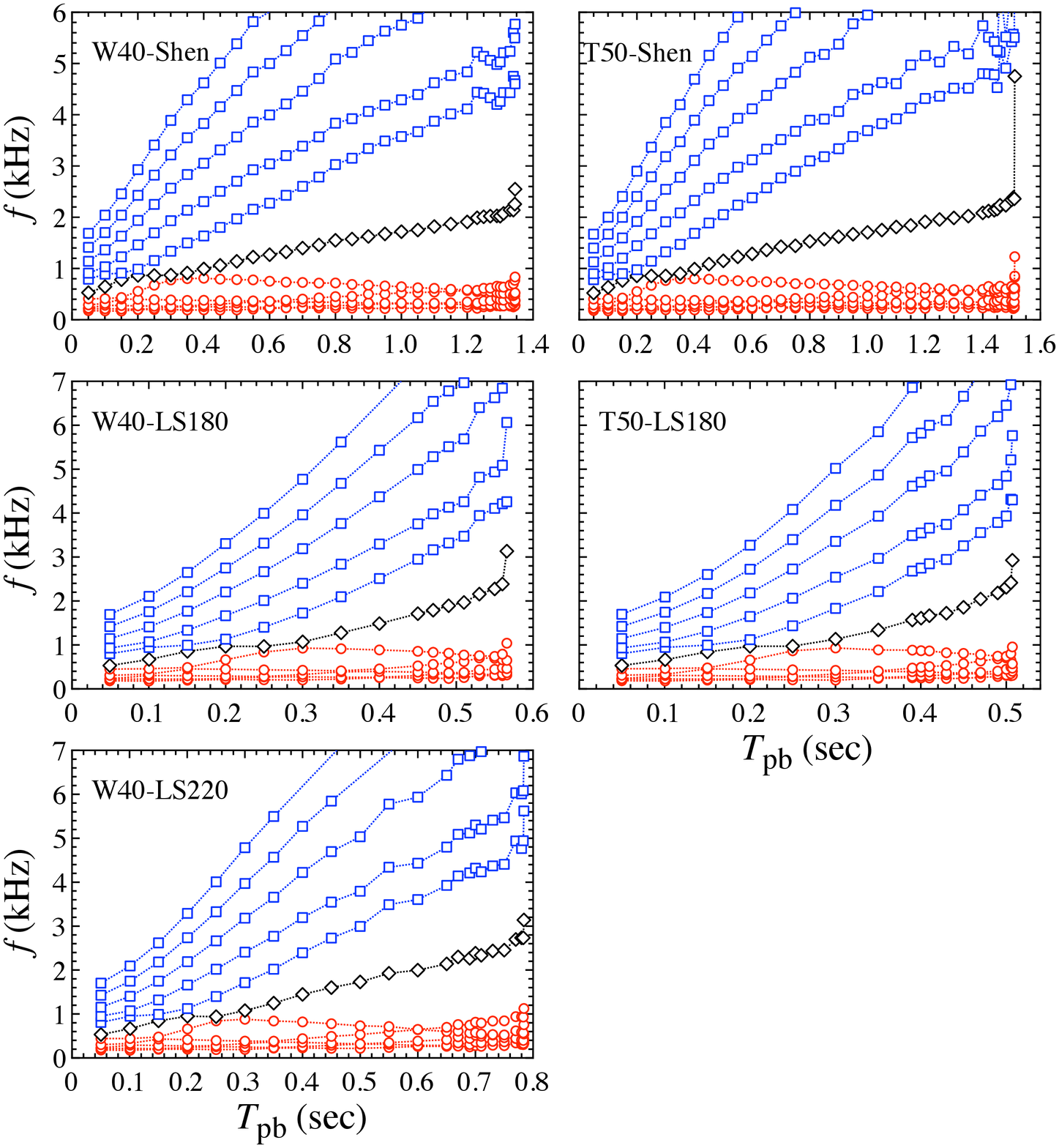} 
\end{center}
\caption{
Time evolution of gravitational wave frequencies from the various PNS models. The left and right panels correspond to the results with W40 and T50, while the top, middle, and bottom panels correspond to the results from the PNS constructed with Shen, LS180, and LS220, respectively. In all panels, the $f$-, $g_i$-, and $p_i$-modes with $i=1-5$ are shown.
Note that we adopt a small time interval for the snapshots for the final phase.  
}
\label{fig:GW}
\end{figure}

We discuss the evolution of gravitational wave frequencies with the PNS model of W40-Shen in main text, as shown in Fig. \ref{fig:W40Shen}. In this appendix, we show the evolution of gravitational wave frequencies from the other PNS models in Fig. \ref{fig:GW}. We remark that the top-left panel is the same as the left panel in Fig. \ref{fig:W40Shen}.
We cover the time periods up to the black hole formation (the formation of apparent horizon) with a small time interval for the final phase.

\end{document}